\documentclass[%
reprint,
superscriptaddress,
 amsmath,amssymb,
 aps,
]{revtex4-1}

\usepackage{graphicx}
\usepackage{dcolumn}
\usepackage{bm}

\begin{document}
\preprint{APS/123-QED}

\title{Evidence of the universal dynamics of rogue waves}

\author{D. Pierangeli}
\affiliation{Dipartimento di Fisica, Universit\`{a} di Roma  ``La Sapienza'', 00185 Rome, Italy}
\email{Davide.Pierangeli@roma1.infn.it}

\author{F. Di Mei}
\affiliation{Dipartimento di Fisica, Universit\`{a} di Roma  ``La Sapienza'', 00185 Rome, Italy}
\affiliation{Center for Life Nano Science@Sapienza, Istituto Italiano di Tecnologia, 00161 Rome, Italy}

\author{C. Conti}
\affiliation{Dipartimento di Fisica, Universit\`{a} di Roma  ``La Sapienza'', 00185 Rome, Italy}
\affiliation{ISC-CNR, Universit\`{a} di Roma ``La Sapienza", 00185 Rome, Italy}

\author{E. DelRe}
\affiliation{Dipartimento di Fisica, Universit\`{a} di Roma  ``La Sapienza'', 00185 Rome, Italy}

\date{\today}

\vspace*{0.2cm}
\begin{abstract} \noindent Light manifests extreme localized waves with long-tail statistics that seem analogous to the still little understood rogue waves in
oceans, and optical setups promise to become laboratory test-beds for their investigation. However, to date there is no evidence that optical extreme events share the
dynamics of their oceanic counterparts, and this greatly limits our ability to study rogue wave predictability using light. Using the Grassberger-Procaccia embedding
method, we here demonstrate that optical spatial rogue wave data in photorefractive crystals has the same predictability and dynamic features of ocean rogue waves.
For scales up to the autocorrelation length, a chaotic and predictable behavior emerges, whereas complexity in the dynamics causes long-range predictability to be limited by the finite size of data sets. The appearance of same dynamics validates the conjecture that rogue waves share universal features across different physical systems,
these including their predictability. 

\begin{description}
\item[PACS numbers] 
\verb+42.65.Sf+, \verb+05.45.-a+ , \verb+05.45.Yv+ 
\end{description}
\end{abstract}
\pacs{Valid PACS appear here}

\maketitle 

\section{Introduction}

Rogue waves are dangerously high water waves that still defy our understanding. Analogous giant perturbations in very different fields of physics, from earthquake, financial and network dynamics to linear and nonlinear optical propagation, appear to share many common features \cite{Onorato2013,
Dudley2014, Zhen-Ya2010, Clauset2009, Solli2007, Lecaplain2012, Montina2009, Oppo2013, Birkholz2013}. A basic challenge is to provide direct evidence of this universality. In Optics, for example, extreme waves have the same statistics of ocean rogue waves but, to date, no evidence of a universality in the underlying dynamics has been reported.  Universal traits in the dynamics are especially important since they would indicate a common predictability of the extreme events.  In most optical cases, the formation of rogue events is accompanied by the onset of noise-seeded instabilities and turbulent dynamics \cite{Walczak2015, Hammani2010, Genty2010} 
so that disorder and non-equilibrium seem common key ingredients triggering long-tail statistics \cite{Armaroli2015, Arecchi2011, Liu2015}. 
Open to debate is the role of nonlinearity. Extreme spatio-temporal events have been observed in linear optical cavities for random fields presenting spectral
inhomogeneity \cite{Arecchi2011, Liu2015, Hohmann2010, Leonetti2015, Dudley2015} as well as in highly-nonlinear regimes in optical crystals, as in Kerr and 
photorefractive feedback systems \cite{Louvergneaux2013, Marsal2010, Marsal2014}. 
In spite of the variety of these observations, the universality between ocean and optical rogue waves remains based only on their anomalous probability distribution 
and on their qualitative waveforms. In fact, Birkholz et al. \cite{Birkholz2015, Erkintalo2015} have recently shown through nonlinear time series analysis how the
dynamical features, such as the chaotic nature and the predictability of the process, are generally different. The inspected optical systems are shown to support rogue
waves through a time process that appear either completely stochastic or completely deterministic, whereas the ocean dynamics presents small-scale predictable traits
and large-scale unpredictability \cite{Birkholz2015}.

In this work we provide the first evidence of optical rogue waves with the same dynamics and predictability of ocean rogue waves. We examine extreme events that
occur in photorefractive beam propagation in turbulent ferroelectrics \cite{Pierangeli2015}. Through the application of the Grassberger-Procaccia
embedding method in the spatial domain \cite{Abarbanel1993, Vulpiani2010}, we find spatial chaos and significant predictability on scales up to the series autocorrelation
length and we identify the complexity of such chaotic dynamics as a key ingredient in determining the observed large-scale unpredictability.

\section{Data and methods}
In our analysis we consider the data series shown in Fig. 1(a), the output intensity distribution of light propagating in an out-of-equilibrium photorefractive 
ferroelectric in conditions leading to spatial rogue waves. This occurs when a Kerr-saturated highly-nonlinear propagation is also affected by instabilities and a
disordered response \cite{Pierangeli2015}. This series, containing a total number of samples $N\simeq2.5 \times 10^4$, represents detected intensities as a function
of the free spatial coordinate $x$ and exhibits a marked long-tail statistics (Fig. 1(d)). In Fig. 1(b) we report a sub-segment of the experimental spatial series
(sampled along the white-dotted line) that includes a rogue wave with its extreme intensity spot. The autocorrelation function for this observation is shown in Fig. 1(c),
with the averaged autocorrelation function obtained considering the whole series of $N$ points. The autocorrelation length of the series $\ell$ represents a typical 
spatial scale of the disordered intensity distribution and is $\ell\simeq9.5 \mu m$, more than one order of magnitude greater than the experimental resolution.

\begin{figure}[t!]
\begin{center}
\vspace*{-0.3cm}
\hspace*{-0.4cm} 
\includegraphics[width=1.07\columnwidth]{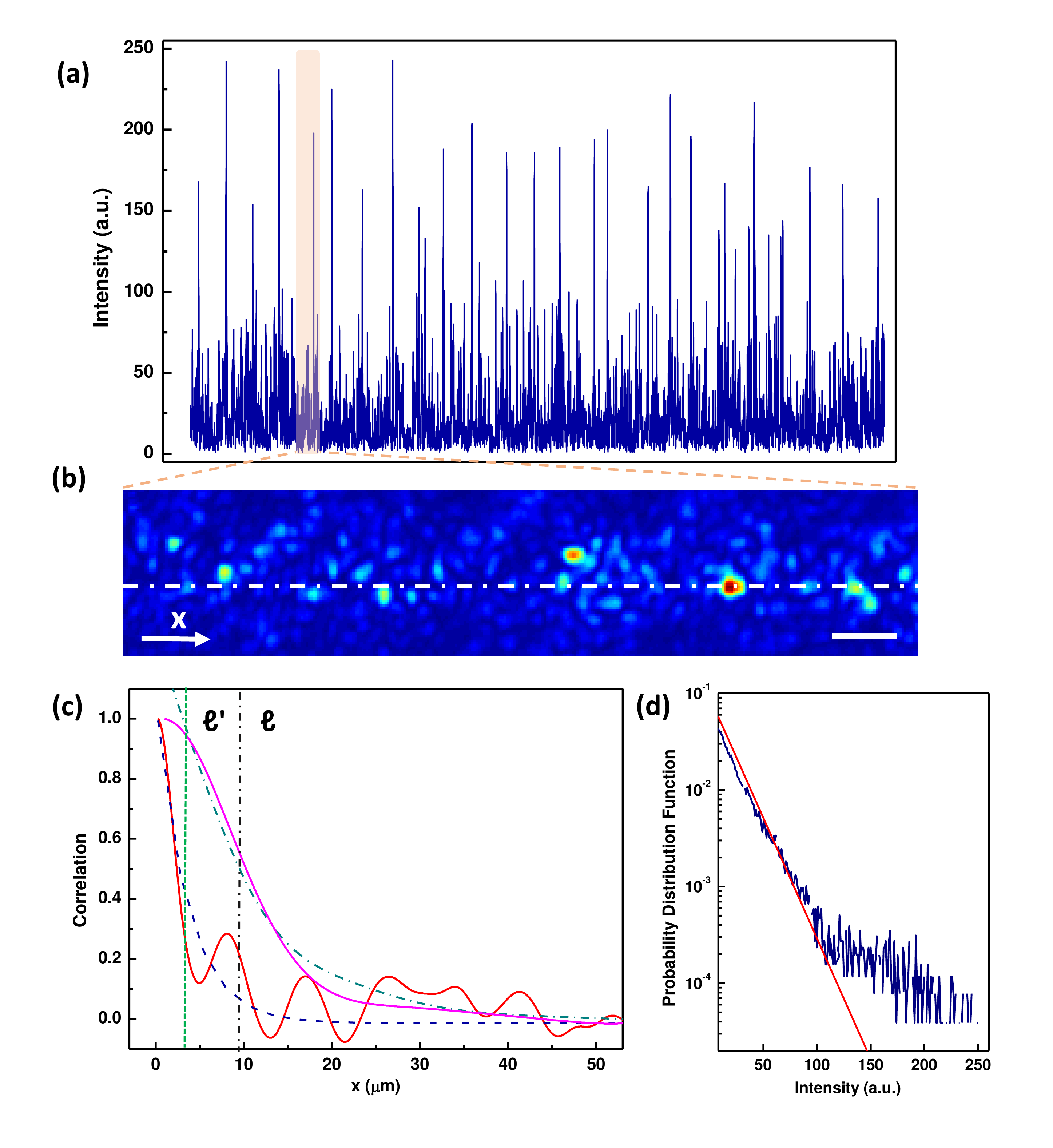} 
\vspace*{-0.5cm}
\caption{Spatial series of intensities detected in a photorefractive ferroelectric with rogue events. (a) Full data series of $N$ samples with the
shaded region underlining the observation shown in (b) and containing a rogue event as an anomalously-bright localized spot. (c) Spatial
autocorrelation functions for the whole series in (a), magenta line, and for the sub-segment shown in (b), red line. Dotted lines in (c) are decay fits giving 
respectively the correlation lengths ($1/e$ width) $\ell$ and $\ell^\prime$. (d) Long-tail statistics of the series (blue line) compared with a normal probability
distribution with the same spatial-averaged intensity (red line).
} 
\end{center}
\label{Figure1}
\end{figure}

The presence of chaotic and predictable features in the spatial intensity distribution defining photorefractive rogue waves is explored with the Grassberger-Procaccia
embedding method \cite{Birkholz2015, Abarbanel1993, Vulpiani2010, Grassberger1983}; specifically, from the $N$-point series $\mathbf{x}=\{ x_1,x_2,...,x_N \}$ (Fig. 1(a))
we consider all the subseries $\mathbf{x}_{im}=\{ x_i,x_{i+1},...,x_{i+m}\}$ of dimension $m$ (embedding dimension).
The statistical distance
\begin{equation}
r_{ijm}=\sqrt{ \sum_{k=i}^{i+m} |x_k - x_{k+j-i}|^2},
\end{equation}
that quantifies the difference between values assumed in two generic subseries, is evaluated for all $i$ and $j>i$ to compute the correlation integral $C_m(r)$ as: 
\begin{equation}
C_m(r)= \frac{2| {r_{ijm}:  (r+\delta r) < r_{ijm}\leq r |}}{(N-m)(N-m-1)}.
\end{equation}
The behavior of $C_m(r)$ at small $r$ reflects the possible appearance of specific subseries with high (non-random) frequency, i.e., it is sensitive to possible
``deja-vu'' phenomena when the series are sampled with a given scale $m$. More specifically, a $C_m(r)\propto r^\nu$ behavior at small $r$ is predicted,
with an exponent $\nu$ that, for large embedding dimensions $m$, characterizes the fractal correlation dimension of the attractor $D(2)$ describing the possible
chaotic dynamics \cite{Vulpiani2010, Grassberger1983}. 
To extract the predictable or stochastic features in our series, we improve the embedding analysis with the method of surrogates, that, as shown in Ref. \cite{Birkholz2015}
for temporal ocean and optical rogue events, allows the comparison of the detected dynamics with the corresponding dynamics that would emerge from a pure random process. 
Starting from physical random data, we compile surrogate data sets identical to the original series as regards for the linear statistical properties. These surrogates have
the same probability distribution function (long-tail statistics), autocorrelation functions and Fourier spectrum of our original series, and appear as reordered copies
of the observed spatial sequence $\mathbf{x}$ \cite{Theiler1992, Schreiber1996, Schreiber1998}.

\section{Same predictability of optical and ocean rogue waves}

Evaluating $C_m(r)$ with surrogate data sets and comparing it with results for the original series we expect small differences if our data comes from a stochastic process.
In contrast, as shown in Fig. 2(a) for $m=12$, large deviations are observed at low $r$, with the rogue series that leads to enhanced recurrence frequencies compared to
the average ones in the surrogates. The observed intensities with extreme events are therefore in principle predictable, meaning that the optical field in a
point of space is directly related to that existing at least at distances smaller than $m$. We quantify this predictability giving the significance of the discrepancy 
between $C_m(r)$ for original and surrogate data; this is reported in Fig. 2(b) (blue line) in units of surrogate standard deviations $\sigma_s$ and has been evaluated
as an average difference at low $r$.
Large predictability characterizes the series on small and intermediate scales $m$, then it vanishes as the autocorrelation length $\ell$ is reached,
in proximity of $m=42$ (corresponding to $10 \mu m$). In other words, on larger scales the spatial dynamics of photorefractive rogue waves lose traces of determinism 
and appear indistinguishable from a re-ordered random series in which the same values of intensity appears.

This scale-dependent behavior is at odds with other reported optical data in the temporal domain, that is either wholly predictable for rogue events in optical filamentation 
or wholly random for those observed in optical fibers \cite{Birkholz2015}. In turn, it remarkably mimics the features of the temporal dynamics of ocean rogue waves.
To see this we extract from Ref. \cite{Birkholz2015} the corresponding predictability for ocean data (Draupner data sets, 1995 \cite{Haver1995, Haver1998}) 
and we plot their behavior in Fig. 2(b) (red lines). As observed in our spatial series, ocean rogue events have a chaotic and predictable dynamics for scales up to the
autocorrelation time $\tau$ that turn unpredictable for long time-scales.  In other words, we find that photorefractive spatial rogue waves and oceanic rogue waves share
the same dynamical process, that is, they share a common universal predictability.

\begin{figure}[t!]
\begin{center}
\vspace*{-0.2cm}
\includegraphics[width=0.92\columnwidth]{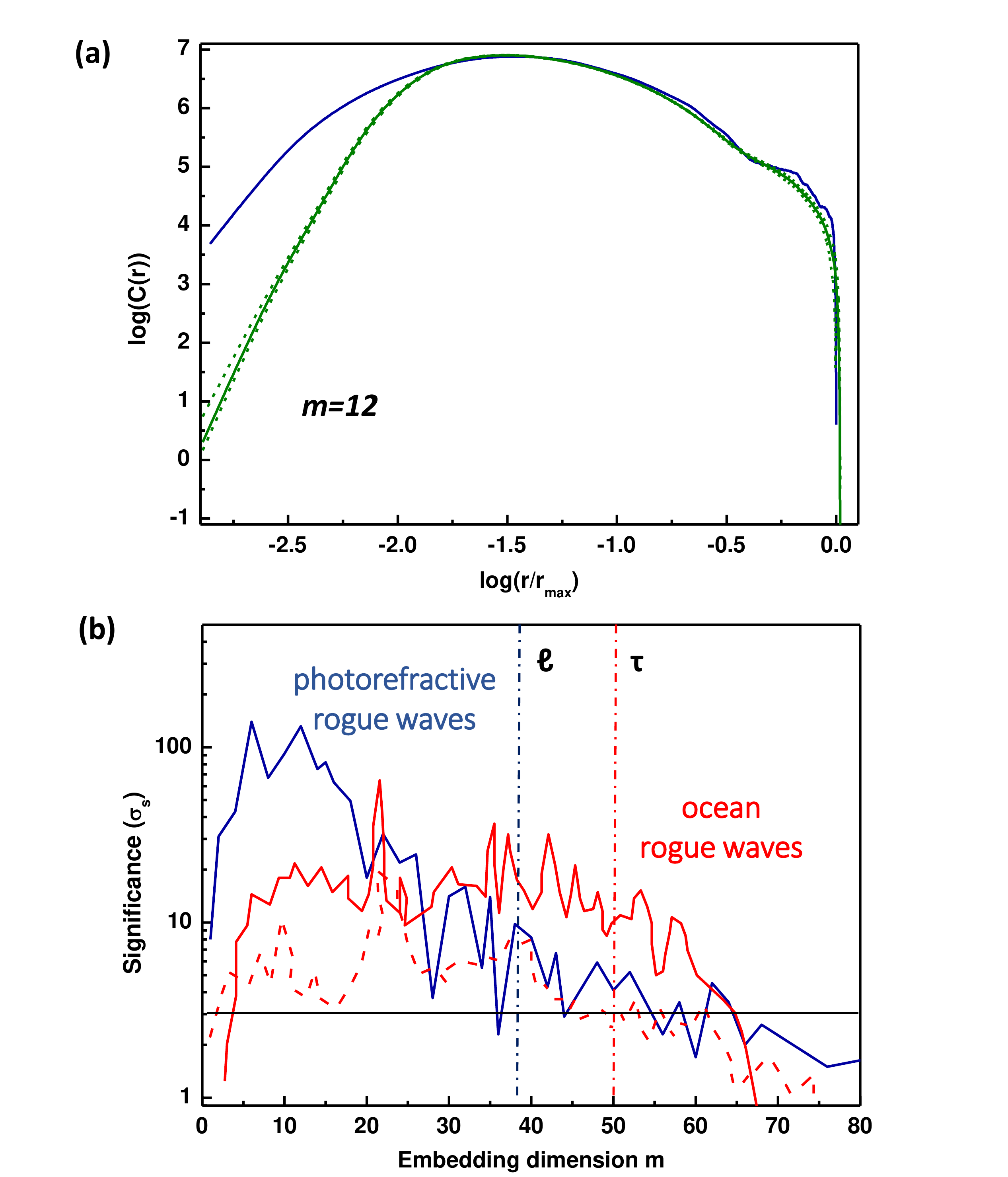} 
\vspace*{+0.2cm}
\caption{Universal dynamics and predictability of rogue waves. (a) $C_m(r)$ at $m=12$ for photorefractive extreme events, showing a large deviation of the rogue 
waves data (blue line) from the average of surrogates (green line), whose spread is indicated by the standard deviation over fifty independent realizations of
surrogates (green-dotted lines). (b) Significance of deviations from surrogates as a function of $m$ for optical data (blue line) and for the temporal dynamics 
of ocean rogue waves (red and red-dotted lines stays respectively for Draupner2 and Draupner1 data, as reported in Ref. \cite{Birkholz2015}). The horizontal black
line indicates the minimum confidences of $3\sigma_s$ separating predictable and unpredictable dynamics; $\ell$ and $\tau$ are respectively the autocorrelation length
and time for the optical and ocean series expressed in term of the embedding dimension $m$.
} \vspace{-0.2cm}
\end{center}
\label{Figure2}
\end{figure}

\begin{figure*}[t!]
\begin{center}
\vspace*{-0.4cm}
\hspace*{-0.4cm}\includegraphics[width=2.15\columnwidth]{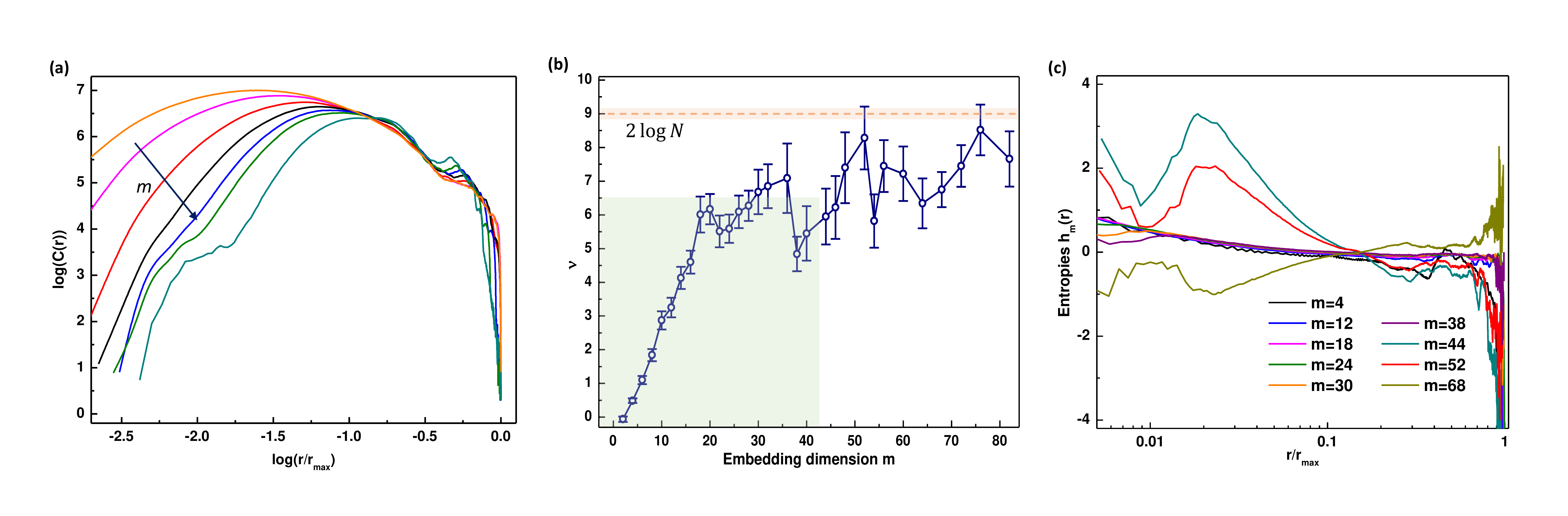} 
\vspace*{-0.5cm}
\caption{Complex chaotic dynamics, predictability, and unpredictability. (a) Correlation integral $C_m(r)$ and (b) its Grassberger-Procaccia exponent $\nu$ varying 
the embedding dimension $m$. (c) Generalized entropies $h_m(r)$ at different $m$, showing the transition at $m \simeq 42$ from the chaotic and predictable behavior
to the unpredictable, unresolved one. The red-shaded line in (b) indicates the maximum observable value of $\nu$ following Ref. \cite{Ruelle1990} for the total
length $N$ of our series. The green-shaded area in (b) indicates meaningful results for which noise and finite-size effects weakly affect the embedding analysis.
} \vspace{-0.2cm}
\end{center}
\label{Figure3}
\end{figure*}

\section{Chaotic and complex dynamics}

A fundamental question arising from the universal behavior shown in Fig. 2(b) is the long-scale unpredictability. In fact, it may be due to an intrinsic property, 
that is, rogue data are actually stochastic on these scales and turbulence rules only locally the spatial dynamics. On the other hand, it may be a ``practical''
unpredictability related to the use of the embedding method for such dynamics. To address the issue we consider $C_m(r)$ for photorefractive rogue waves that, as
reported in log-scale in Fig. 3(a), show a linear behavior at low $r$ with a slope that increases with the embedding dimensions $m$. The extracted exponent $\nu$ as a
function of $m$ is reported in Fig. 3(b). The behavior is characterized by a well-defined linear growth at low $m$, as typically occurs in dynamical processes both
of chaotic and stochastic origin \cite{Vulpiani2010}, since, basically, similarities in subseries are much less likely increasing the subseries size. 
In general, chaos is highlighted by a saturation of $\nu$ at large $m$ and the approached value corresponds to the fractal correlation dimension of the attractor $D(2)$.
We observe that starting from $m=20$ the linear behavior is damped and seems to saturate to a constant value in proximity of $\nu\approx8$.
However, a careful analysis is needed in deducing the chaotic dynamics from Fig. 3(b). Noise in experimental data, emerging from the physical realization of the
process and from its detection, affects especially the estimate of $\nu$ for long subseries, that consequently present a larger uncertainty. More importantly,
a crucial role is played by the amount of available data $N$ \cite{Smith1988, Ruelle1990}. Because the number of independent subseries $\mathbf{x}_{im}$ used in the
embedding method scales as $N/m$, the information that can be extracted is limited when considering long scales. For a series of $N$ points, numerical criteria set 
the maximum observable value of $\nu$ approximately at $2logN$ \cite{Ruelle1990} (horizontal line in Fig. 3(b)), implying that only values sufficiently below this
threshold are reliable. From this point of view, the reported saturation of $\nu$ cannot be used to demonstrate the presence of a low-dimensional attractor.
However, it implies at least the presence of a high-dimensional attractor that underlies the complexity of the chaotic dynamics.
We analyze the scale dependence of the this dynamics using the generalized entropies 
\begin{equation}
h_m(r)= ln \left[ C_m(r) /C_{m+1}(r) \right],
\end{equation}
in which a plateau to a constant value indicates the chaotic origin of the signal \cite{Vulpiani2010}. As shown in Fig. 3(c) flat entropies characterize
the series for $m\lesssim42$, confirming that the signal is revealed as deterministic on these scales. For $m\gtrsim 42$, $h_m(r)$ exhibits a noisy and non-monotonic 
behavior, consistent with the absence of predictable traits, as previously pointed out in Fig. 2(b), and with a complex and non-random dynamics. In fact, stochastic 
series generally presents a linear decreasing of $h_m(r)$ as a function of $r$.
Following these results and the above criterion limiting the detection of $\nu$, we estimate that only results for $m\lesssim 42$ (green area in Fig. 3(b)) are meaningful
to the correlation features shown by the series and, here, only signs of a saturation in $\nu$ are present.
Above this ``transition'', to resolve the long-range dynamics up to $m$, the Grassberger-Procaccia method needs an amount of data $N$ that grows exponentially with
$m$ \cite{Vulpiani2010, Smith1988}. Therefore the large-scale unpredictability for photorefractive rogue waves has roots in the complex behavior of the dynamical process
and the same fact is expected to hold for the ocean ones.

\section{Discussion}

We note that in adressing universality we are here comparing different dynamics in space and time, where the role of nonlinearity is also expected to be greatly different.
Specifically, in the emergence of such universal dynamics a basic tassel may be provided by the generalized nonlinear Schr\"{o}dinger model that is involved in the
description of both cases \cite{Onorato2013, Dudley2014, Pierangeli2015}. However, for water waves it is not established if deterministic solitonic solutions of 
this model \cite{Osborne2000, Baronio2012, Chabchoub2012, Toffoli2015} can describe the observed ocean statistics and dynamics, or if linear interacting random waves can 
also play a key rule in these properties \cite{Liu2015, Stansell2004, Chrisou2011}. For propagation in photorefractive media, optical long-tail statistics emerge in
highly-nonlinear and randomic conditions dominated by inelastic soliton interaction \cite{Pierangeli2015}. This suggests that chaotic traits may lie at the heart of
extreme events independently of the strength of the nonlinear mechanism.

\section{Conclusions}

It has long been speculated that the emergence of rogue waves is characterized by universal properties that can be found across different fields of physics.
Studying the experimental spatial distribution of photorefractive rogue waves and mapping the detected intensity distribution into a dynamical system, 
using embedding methods to analyze recurrences in the series, we have found a chaotic and predictable behavior on scales up to the autocorrelation length,
a physical dynamic that turns out to be the same as that exhibited by ocean rogue waves. This universality is quantitative and completes the analogy between ocean and 
optical extreme events previously based qualitatively on the probability distribution function and on the waveforms. It opens the key possibility to address the
puzzle of ocean rogue wave predictability using light. Indeed, the unpredictability on large scales appears for optical events as a property not related to 
a stochastic process but to the complexity of this dynamics, that, given the finite size of the sample, cannot be resolved to arbitrary distances with the embedding method.
Our results shed light on the universality of extreme events in optics and ocean dynamics and point out chaos and complexity as fundamental physical ingredients
for rogue waves and their predictability in different spatially-extended systems.

\section*{Acknowledgements}

Funding from grants PRIN 2012BFNWZ2, and Sapienza 2014 and 2015 Research grants are acknowledged.
We sincerely thank Prof. G. Steinmeyer and Prof. A. Vulpiani for useful discussions and suggestions.

\end{document}